\documentclass[12pt,eqsecnum]{JHEP3}

\newcommand{\be}{\begin{equation}}
\newcommand{\ee}{\end{equation}}
\newcommand{\bea}{\begin{eqnarray}}
\newcommand{\eea}{\end{eqnarray}}
\newcommand{\nn}{\nonumber}
\newcommand{\n}{\nabla}
\newcommand{\rd}{\partial}

\title{K\"{a}hler Quantization of $H^3(CY_3,R)$ and the Holomorphic Anomaly}
\author{Farhang Loran\\
  Department of  Physics, Isfahan University of Technology (IUT)\\
  Isfahan,  Iran\\
  E-mail: \email{loran@cc.iut.ac.ir}}

 \abstract{Studying the quadratic field theory on seven dimensional spacetime
 constructed by a direct product of Calabi-Yau three-fold by  a real
 time axis, with phase space being the third cohomology of the Calabi-Yau three-fold,
 the generators of translation along moduli directions of
 Calabi-Yau three-fold are constructed. The algebra of these
 generators is derived which take a simple form in
 canonical coordinates. Applying the Dirac method of quantization of second class constraint systems,
 we show that the Schr\"{o}dinger equations corresponding to these generators are equivalent to the holomorphic anomaly
 equations if one defines the action functional of the quadratic
 field theory with a proper factor one-half.}
\begin{document}
 \section{Introduction and Summary}
 In \cite{Vafa,Vafa1}, Bershadsky, Cecotti, Ouguri and Vafa described a holomorphic anomaly in topological string theories obtained by
 twisting $N=2$ models. They described this anomaly as a subtle breakdown of the BRST invariance in the twisted N=2 model coupled to gravity.
 In \cite{Witten} Witten discussed the implication of the
 holomorphic anomaly to the background independence of the string
 theory. He showed that the holomorphic anomaly can be understood as a violation of the naive background independence that explains some general puzzles
 in mirror symmetry as a relation between A-model and B-model of Calabi-Yau manifolds.
 He also derived two equations resembling the holomorphic anomaly
 equations derived in \cite{Vafa} by quantizing $H^3(CY_3,R)$, see also \cite{Verlinde,Gerasimov}. This
 observation is one of the most straightforward realizations of the
 idea that the partition function of the type B topological strings
 on Calabi-yau manifolds is related to a holomorphic wave function
 in  the seven dimensional theory with a phase space being the
 $H^3(CY_3,R)$ \cite{Verlinde}.
 \par
 In this paper we consider  the quadratic field theory on seven dimensional spacetime
 $CY_3\times R$, given by the action,
 \be
 S(C)=\mbox{const.}\int_{CY_3\times R}C\wedge dC,
 \label{ac1}
 \ee
 The phase space of this theory is known to be $H^3(CY_3,R)$ \cite{Verlinde,Gerasimov}.
 We derive the explicit form of $H_i$ and $\bar H_{\bar i}$,  the generators  of translation along moduli directions $t_i$ and
 $\bar t_{\bar i}$ of the  Calabi-Yau three-fold. We show that in canonical coordinates where $\bar H_{\bar i}=\overline{H_i}$, these generators
 satisfy the algebra,
 \bea
 \{\bar H_{\bar i},\bar H_{\bar j}\}&=&0,\nn\\
 \{H_i,H_j\}&=&0,\nn\\
 \{H_i,\bar H_{\bar i}\}&=&\rd_i\bar H_{\bar i}-\bar\rd_{\bar i}
  H_i.
 \eea
 After quantization it is shown that, if  the constant factor in Eq.(\ref{ac1}) is equal to one-half, i.e. for the
 action,
 \be
 S(C)=\frac{1}{2}\int_{CY_3\times R}C\wedge dC,
 \label{trueaction}
 \ee
 the Schr\"{o}dinger equations
 \be
 \begin{array}{ccc}
 i\rd_i\left|\psi\right>=H_i\left|\psi\right>,&&
 i\bar\rd_{\bar i}\left|\psi\right>=\bar H_{\bar
 i}\left|\psi\right>,
 \end{array}
 \ee
 are equivalent to holomorphic anomaly equations in the simplified form given in
 \cite{Verlinde}.

 The organization of the paper is as follows. In section \ref{sec 2},
 the classical theory on $CY_3\times R$ is studied. The generators
 $H_i$ and $\bar H_{\bar i}$ are constructed and their algebra is studied. In section \ref{sec 3},
 it is shown that the Schr\"{o}dinger equations with respect to generators $H_i$ and $\bar H_{\bar i}$
 are equivalent to the holomorphic anomaly equations in topological string theory.
 Finally we determine the true constant factor in definition of the action functional. For this purpose we first study the
 constraint structure of the theory and derive Dirac brackets in section \ref{sec 3A}. Using these results the factor one-half in
 Eq.(\ref{trueaction}) is derived in section \ref{sec 3B}.
 \section{The Calabi-Yau Moduli Space}\label{sec 2}
 A Calabi-Yau manifold ($CY$) is a compact K\"{a}hler manifold with
 vanishing  first Chern class. According to a conjecture by Calabi,
 proven  by Yau, $CY$ manifolds admit a Ricci-flat metric. Thus $CY$ manifolds provide  solutions to the Einstein equation.
 Calabi-Yau three-fold ($CY_3$) is considered for superstring compactification.

 The vanishing  of the first Chern class implies that the canonical bundle is trivial. Thus in the case of $CY_3$, the holomorphic three-form
 $\Omega\in H^{3,0}(CY_3,R)$ (the $CY$-form) is unique only up to a scale. By deformations of the complex structure, a $(3,0)$-form can only
 change into a linear combination of $(3,0)$-form and
 $(2,1)$-forms, thus $H^{2,1}$ measures complex structure deformation.
 In fact, $H^3$ forms a bundle over the moduli space of complex
 structures of $CY_3$ and the $CY$-form $\Omega$  defines a line sub-bundle $\cal L$. A K\"{a}hler potential can be defined by
 the relation
 \be
 K=-\ln i\int \bar\Omega\wedge\Omega,
 \label{K-potential}
 \ee
 which transforms under $\Omega\to e^{f(z)}\Omega$ as $K\to K-f-\bar
 f$. On the line bundle $\cal L$ one can define a connection
 $\n_i=\rd_i+\rd_iK$ which curvature $G_{ij}=\rd_i\bar\rd_{\bar j}K$
 is the K\"{a}hler metric on the moduli space. It is a simple exercise to show that
 $\n_i\Omega$ is the $H^{2,1}$ part of $\rd_i\Omega$.

 A real closed $\gamma\in H^3(CY_3,R)$ can be decomposed in the basis
 consisting the holomorphic three-form $\Omega\in H^{3,0}$, its
 covariant derivatives $\n_i \Omega\in H^{2,1}$ and their
 complex conjugates,
 \be
 \gamma=\lambda^{-1}\Omega+x^i\n_i\Omega+\bar x^{\bar i}\bar\n_{\bar
 i}\bar\Omega+\bar\lambda^{-1}\bar\Omega.
 \label{gamma}
 \ee
 $\gamma$ can be considered as the classical solution of the
 Euler-Lagrange equation of motion in the seven dimensional field
 theory on $CY_3\times R$ with action functional \cite{Verlinde,Gerasimov},
 \be
 S(C)=\int_{CY_3\times R}C\wedge dC,
 \label{action}
 \ee
 where $C$ is a real three-form. Using the decomposition
 $C=\gamma+\omega dt$ in terms of $\gamma$ a three-form component of
 $C$ along $CY_3$ and $\omega dt$ a two-form along $CY_3$ and a one
 form along $R$, the action can be written as,
 \be
 S(\gamma,\omega)=\int dt\int_{CY_3}\left(\gamma\frac{\rd}{\rd t}\gamma+\omega
 d\gamma\right),
 \label{action1}
 \ee
 which equation of motion imply that $\gamma$ is closed and time
 independent, see section \ref{sec 3A}. Thus the action functional (\ref{action}) can be used to quantize $H^3(CY_3)$.  The phase space can be constructed using the
 coordinates $\lambda^{-1},x^i$'s defined in Eq.(\ref{gamma}) and their
 conjugate momenta defined by the action (\ref{action}),
 \be
 p_i=ie^{-K}G_{i\bar i}\bar x^{\bar i}, \hspace{1cm} \pi=-ie^{-K}\bar\lambda^{-1},
 \label{momenta}
 \ee
 with Poisson brackets,
 \be
 \{x^i,p_j\}=\delta^i_j,\hspace{1cm} \{\lambda^{-1},\pi\}=1,
 \label{poisson}
 \ee
  with all other Poisson brackets vanishing.
 We assume that the phase space operators $x^i$, $p_j$, $\lambda^{-1}$ and $\pi$ do not explicitly depend on the moduli $t\bar
 t$.   Defining $H_i$ and $\bar H_{\bar i}$ to be generators of translation along moduli
 directions   $t_i$ and $\bar t_{\bar i}$ respectively,  one has,
 \bea
 {\cal O}_{,i}&=&\{{\cal O},H_i\},\nn\\
 {\cal O}_{,\bar i}&=&\{{\cal O},\bar H_{\bar i}\}.
 \label{Heisenberg}
 \eea
 We use the convention ${\cal O}_{,i}=\partial_i{\cal O}=\frac{\partial}{\partial t_i}{\cal
 O}$ occasionally.  To obtain $H_i$ and $\bar H_{\bar i}$ we note that $\gamma$ defined
 in Eq.(\ref{gamma}) is independent of moduli. For example,
 \bea
 \gamma_{,\bar i}&=&\lambda^{-1}_{,\bar i}\Omega+x^i_{,\bar
 i}\n_i\Omega+x^i\bar\rd_{\bar i}\n_i \Omega+\bar x^{\bar j}_{,\bar
 i}\bar\n_{\bar j}\bar\Omega+{\bar x }^{\bar j}\bar\rd_{\bar
 i}\bar\n_{\bar j}\bar\Omega+\bar\lambda^{-1}_{,\bar
 i}\bar\Omega+\bar\lambda^{-1}\bar\rd_{\bar i}\bar\Omega\nn\\
 &=&(\lambda^{-1}_{,\bar i}+x^iG_{i\bar i})\Omega+(x_{,\bar i}^k-e^K\bar
 x^{\bar j}G^{k\bar k}\bar C_{\bar i \bar j \bar k})\n_k\Omega+(\bar
 x^{\bar k}_{,\bar i}+\bar\lambda^{-1}\delta^{\bar k}_{\bar
 i}-\bar\rd_{\bar i}K-\bar x^{\bar j}\Gamma^{\bar k}_{\bar i\bar
 j})\bar\n_{\bar k}\bar\Omega\nn\\
 &+&(\bar\lambda^{-1}_{,\bar
 i}-\bar\lambda^{-1}\bar\rd_{\bar i}K)\bar\Omega\nn\\
 &=&0.
 \label{b}
 \eea
 To obtain the second equality above we have used the identities
 (\ref{a2}) that can be verified using the following equations,
 \bea
 e^{-K}&=&i\int_{CY_3}\bar\Omega\wedge\Omega,\nn\\
 e^{-K}G_{i\bar j}&=&i\int_{CY_3}\n_i\Omega\wedge\bar\n_{\bar
 j}\bar\Omega,\nn\\
 C_{ijk}&=&i\int_{CY_3}\n_i\Omega\wedge D_i\n_k\Omega,
 \label{a1}
 \eea
 where $C_{ijk}$ is the three point function in the B-model,
 $C_{ijk}=-i\int_{CY_3}\Omega\wedge\rd_i\rd_j\rd_k\Omega$.
 The covariant derivative $D_i$ in Eq.(\ref{a1}) contains the usual
 Christoffel connection $\Gamma^k_{ij}=-G^{k\bar k}G_{k\bar j,{\bar i}}$ as well as the term
 $\rd_iK$ see for example Eq.(\ref{Connection D}). Eq.(\ref{a1}) implies that modulo exact terms one has,
 \be
 \bar\rd_{\bar i}\n_j\Omega=G_{\bar i
 j}\Omega,\hspace{1cm}D_i\nabla_j\Omega=-e^KG^{k\bar
 k}C_{ijk}\bar\n_{\bar k}\bar\Omega.
 \label{a2}
 \ee
 From Eq.(\ref{b}) and its complex conjugate ($\gamma_{,i}=0$) one obtains,
 \be
 \begin{array}{lll}
 \lambda^{-1}_{,\bar i}=-G_{\bar i i}x^i,&&\bar\lambda^{-1}_{,i}=-G_{i\bar
 i}\bar x^{\bar i},\\
 x^k_{,\bar i}=e^K G^{k\bar k}\bar C_{\bar i\bar j \bar k}\bar
 x^{\bar j},&&\bar x^{\bar k}_{,i}=e^KG^{\bar k k}C_{ijk}x^j,\\
 \bar x^{\bar k}_{,\bar i}=-\bar \lambda^{-1}\delta^{\bar k}_{\bar i}+\bar
 x ^{\bar j}\bar\Gamma^{\bar k}_{\bar i\bar j}+\bar\rd_{\bar
 i}K\bar x^{\bar k},&&x^k_{,i}=-\lambda^{-1}\delta^k_i+x^j\Gamma^k_{ij}+\rd_iKx^k,\\
 \bar\lambda^{-1}_{,\bar i}=\bar\lambda^{-1}\bar\rd_{\bar
 i}K,&&\lambda^{-1}_{,i}=\lambda^{-1}\rd_iK.
 \end{array}
 \label{c}
 \ee
 It is straightforward to verify that using Eqs.(\ref{momenta}), (\ref{poisson}) and (\ref{Heisenberg}) one can obtain Eq.(\ref{c}) if
 \bea
 H_i&=&-\lambda^{-1} p_i-\frac{i}{2}C_{ijk}x^jx^k+\frac{\rd_iK}{2}[\pi,\lambda^{-1}]_++\frac{1}{2}\Gamma^k_{ij}[x^j,p_k]_+
 +\frac{\rd_iK}{2}[x^j,p_j]_+,\nn\\
 \bar H_{\bar i}&=&-G_{\bar i i}\pi x^i-\frac{i}{2}e^{2K}\bar
 C_{\bar i\bar j\bar k}G^{\bar j j}G^{\bar k k}p_jp_k,
 \label{Hamiltonian}
 \eea
 where we have used the notation $[A,B]_+=AB+BA$ for later convenience. At the classical
 level $[A,B]_+=2AB$ as far as there is no operator ordering
 problem. For example to obtain the equality $\bar\lambda^{-1}_{,i}=-G_{i\bar
 i}\bar x^{\bar i}$ given in Eq.(\ref{c}), one uses Eqs.(\ref{Heisenberg}) and (\ref{Hamiltonian}) to obtain
 $\pi_{,i}=p_i-K_{,i}\pi$. Then using the relation
 $\pi=-ie^{-K}\bar\lambda^{-1}$ given in Eq.(\ref{momenta}) to calculate $\bar\lambda^{-1}_{,i}$ in terms of $\pi_{,i}$, one
 obtains the desired result.

 The algebra of $H_i$ and $\bar H_{\bar i}$ is interesting. It is
 easy to verify that
 \be
 \{\bar H_{\bar i},\bar H_{\bar j}\}=0.
 \label{HbarHbar commutator}
 \ee
 After some calculations one can show that
 \bea
 \{H_i,H_j\}&=&i\left[\left(C_{jlk}\Gamma^k_{in}+C_{jln}\rd_iK\right)-i\leftrightarrow
 j\right]x^lx^n+\left(\Gamma^k_{im}\Gamma^m_{jl}-i\leftrightarrow j\right)x^lp_k\nn\\
 &=&i\left[(D_i-\rd_i)C_{jnl}-i\leftrightarrow j\right]x^lx^n-\left(\rd_i\Gamma^k_{jl}-i\leftrightarrow j\right)x^lp_k\nn\\
 &=&i\left(\rd_jC_i-\rd_iC_j\right)_{nl}x^lx^n+\left(\rd_j\Gamma^k_{il}-\rd_i\Gamma^k_{jl}\right)x^lp_k.
 \label{H commutator1}
 \eea
 The connection $D_i$ is defined by the relation,
 \be
 D_iC_j=\rd_iC_j+\rd_iKC_j+\Gamma^k_{ij}C_k,
 \label{Connection D}
 \ee
 The second and third equalities in Eq.(\ref{H commutator1}) are  obtained using the fact that the four point function $C_{ijkl}=D_iC_{jkl}$ is
 totally symmetric in its four indices \cite{Vafa}, which is the consequence of the  $tt^*$ equation
 $D_iC_j=D_jC_i$ \cite{CVafa}. We have also used the identity,
 \be
 R^k_{jil}=\left(\rd_i\Gamma^k_{jl}+\Gamma^k_{im}\Gamma^m_{jl}\right)-i\leftrightarrow
 j=0.
 \ee
 Eq.(\ref{H commutator1}) can be more simplified using the WDVV (Witten-Dijkgraaf-Verlinde-Verlinde) equation $\rd_iC_j=\rd
 _jC_i$ to obtain,
 \be
 \{H_i,H_j\}=\rd_jH_i-\rd_iH_j.
 \label{H commutator2}
 \ee
 Finally one can show that
 \bea
 \{H_i,\bar H_{\bar i}\}&=&\rd_i\bar H_{\bar i}-\bar\rd_{\bar i}H_i.
 \label{H Hbar commutator}
 \eea
 To obtain the above equality we have used the identities,
 \bea
 \rd_i\bar C_{\bar i\bar j\bar k}&=&\bar\rd_{\bar i}C_{ijk}=0
 ,\nn\\
 R_{i\bar j k}^l=-\bar\rd_{\bar j}\Gamma^l_{ki}&=&G_{k\bar
 j}\delta^l_i+G_{i\bar j}\delta^l_k-e^{2k}C_{ikn}G^{n\bar n}\bar
 C_{\bar j\bar m \bar n}G^{\bar m l},
 \eea
 which result in,
 \bea
 -\bar\rd_{\bar i} H_i&=&-G_{i\bar i}\left(\lambda^{-1}\pi+x^kp_k\right)+R_{i\bar i
 k}^lx^kp_l,\nn\\
 -\rd_i\bar H_{\bar i}&=&-\pi\Gamma^k_{ij}G_{k\bar
 i}x^j+ie^{2K}G^{j\bar j}G^{l\bar k}\bar C_{\bar i\bar j\bar
 k}\Gamma^l_{ij}p_jp_l+i\rd_iKe^{2K}G^{j\bar j}G^{l\bar k}\bar C_{\bar i\bar j\bar
 k}p_jp_k.
 \eea
  More interesting relations can be obtained in canonical coordinates. In canonical coordinates $\rd_iK$ and $\Gamma^k_{ij}$  together with
  all holomorphic derivatives are vanishing locally.
 Thus in these coordinates,  the right hand side of the first equality in Eq.(\ref{H commutator1})  is simply vanishing.
 Consequently, at least locally,
 \bea
 \{\bar H_{\bar i},\bar H_{\bar j}\}&=&0,\nn\\
 \{H_i,H_j\}&=&0,\nn\\
 \{H_i,\bar H_{\bar i}\}&=&\rd_i\bar H_{\bar i}-\bar\rd_{\bar i}
 H_i.
 \eea
 Furthermore using the definitions (\ref{momenta}) and
 (\ref{Hamiltonian}) one obtains
 \be
 \bar H_{\bar i}=\overline{H_i}.
 \ee
 \section{Quantization and The Holomorphic Anomaly}\label{sec 3}
 For quantization we consider, as usual, the phase space coordinates
 as operators satisfying the commutation relations consistent with
 the Poisson algebra (\ref{poisson}),
 \be
 [x^i,p_j]=i\delta^i_j,\hspace{1cm} [\lambda^{-1},\pi]=i,
 \label{commutator}
 \ee
 with all other commutators vanishing. Quantum states satisfy the
 set of  Schr\"{o}dinger equations,
 \bea
 i\rd_i\left|\psi\right>&=&H_i\left|\psi\right>,\nn\\
 i\bar\rd_{\bar i}\left|\psi\right>&=&\bar H_{\bar
 i}\left|\psi\right>.
 \label{Schrodinger}
 \eea
 Using the commutation relations (\ref{commutator}), the generator $H_i$ defined in  Eq.(\ref{Hamiltonian}) can be
 written in a simple form,
 \be
 H_i=-\lambda^{-1}
 p_i-\frac{i}{2}C_{ijk}x^jx^k-\rd_iK\left(-\lambda^{-1}\pi-x^jp_j+i\frac{h+1}{2}\right)+
 \Gamma^k_{ij}x^jp_k-\frac{i}{2}\ln\left|G\right|_{,i},
 \ee
 where we have used the identity $\ln\left|G\right|_{,i}=\Gamma^j_{ij}$ and $h=h^{2,1}$.
 In $(x,\lambda^{-1})$ space, where $p_i=-i\frac{\rd}{\rd x^i}$ and $\pi=-i\frac{\rd}{\rd \lambda^{-1}}$, the set of Schr\"{o}dinger equations
 (\ref{Schrodinger}) give,
 \bea
 \bar\rd_{\bar i}\Psi&=&\left[G_{\bar i i} x^i\frac{\rd}{\rd\lambda^{-1}}+\frac{1}{2}e^{2K}\bar
 C_{\bar i\bar j\bar k}G^{\bar j j}G^{\bar k k}\frac{\rd^2}{\rd x^j\rd x^k}\right]\Psi,\nn\\
 \left[\n_i+\Gamma^k_{ij}x^j\frac{\rd}{\rd x^k}\right]\Psi&=&\left[\lambda^{-1}\frac{\rd}{\rd
 x^i}-\frac{1}{2}\ln\left|G\right|_{,i}-
 \frac{1}{2}C_{ijk}x^jx^k\right]\Psi,
 \label{holomorphic}
 \eea
 where the connection $\n_i$ is given by
 \be
 \n_i=\rd_i+\rd_iK\left(\frac{h+1}{2}+x^j\frac{\rd}{\rd
 x^j}+\lambda^{-1}\frac{\rd}{\rd\lambda^{-1}}\right).
 \ee
 Eq.(\ref{holomorphic}) is the holomorphic anomaly equation obtained in \cite{Verlinde}.
 \subsection{Constraint structure and Dirac brackets}\label{sec 3A}
 In this section we digress to study the constraint structure of the action
 (\ref{action}) which is essential to obtain the true coefficient of the action functional
 (\ref{action}). For an introduction to constraint systems see
 \cite{Dirac}.

 The second term in Eq.(\ref{action1}) introduces a first class constraint $d\gamma=0$ which makes $\gamma$ closed. This is a
 secondary constraint following the primary
 constraint $\pi_{\omega}=\frac{\delta S}{\delta \dot \omega}=0$ where $\pi_\omega$ denotes the momentum conjugate to the three-form $\omega$.
 Using Eq.(\ref{gamma}) and (\ref{a1}), the first term in the action (\ref{action1}) in terms of coordinates
 $\lambda^{-1}$, $x^i$'s and their complex conjugates can be written as follows,
 \be
 S(\lambda^{-1},x^i,\bar\lambda,\bar\lambda^{-1})=ie^{-K}\left(\lambda^{-1}\dot{\bar\lambda}^{-1}-\dot{\lambda}^{-1}\bar\lambda^{-1}\right)-
 ie^{-K}G_{i\bar i}\left(x^i\dot{\bar x}^{\bar i}-\dot x^i\bar x^{\bar
 i}\right).
 \label{action2}
 \ee
 The momenta conjugate to the coordinates $\lambda^{-1}$, $x^i$'s and their complex
 conjugates can be defined using the general rule,
 \be
 \begin{array}{lll}
 \pi=\frac{\delta S}{\delta \dot\lambda^{-1}}=-i
 e^{-K}\bar\lambda^{-1},&&\bar\pi=\frac{\delta S}{\delta \dot{\bar\lambda}^{-1}}=i
 e^{-K}\lambda^{-1},\\\\
 p_i=\frac{\delta S}{\delta \dot x^i}=ie^{-K}G_{i\bar i}\bar x^{\bar
 i},&&\bar p_{\bar i}=\frac{\delta S}{\delta \dot{\bar x}^{\bar i}}=-ie^{-K}G_{i\bar i}
 x^i.
 \end{array}
 \label{momenta1}
 \ee
 Since momenta are independent of velocities, the above equations
 in fact are relations between phase space coordinates, i.e.  they are Dirac constraints.
 Denoting these constraints as $\chi_I$, $I=1,\cdots,2h+2$,
 \be
 \chi_I=\left\{
 \begin{array}{lll}
 \chi_{i},&&I=1,\cdots,h\\
 \bar\chi_{\bar i},&& I=h+1,\cdots,2h\\
 \chi_0,&&I=2h+1,\\
 \bar\chi_0&&I=2h+2,
 \end{array}\right.
 \ee\
 where,
 \be
 \begin{array}{lll}
 \chi_i=p_i-ie^{-K}G_{i\bar i}\bar x^{\bar i},&&\bar\chi_{\bar i}=\bar
 p_{\bar i}+ie^{-K}G_{i\bar i} x^i,\\\\
 \chi_{0}=\pi+i e^{-K}\bar\lambda^{-1},&&\bar\chi_{0}=\bar\pi-i
 e^{-K}\lambda^{-1},
 \end{array}
 \label{constraints}
 \ee
 are constraints defined by Eq.(\ref{momenta1})  and using the canonical Poisson brackets,
 \be
 \begin{array}{ccc}
 \{x^i,p_j\}=\delta^i_j,&&\{\bar x^{\bar i},\bar p_{\bar
 j}\}=\delta^{\bar i}_{\bar
 j},\\\{\lambda^{-1},\pi\}=1&&\{\bar\lambda^{-1},\bar\pi\}=1,
 \end{array}
 \label{Poisson}
 \ee
 with all other Poisson brackets vanishing, one verifies that
 \bea
 \{\chi_i,\bar\chi_{\bar i}\}&=&-2ie^{-K}G_{i\bar i},\nn\\
 \{\chi_0,\bar\chi_{0}\}&=&2ie^{-K},\nn\\
 \{\chi_0(\bar\chi_{0}),\chi_i\}&=&0,\nn\\
 \{\chi_0(\bar\chi_{0}),\bar\chi_{\bar i}\}&=&0.
 \eea
 Therefore,
 \be
 \det\left(\{\chi_I,\chi_J\}\right)\neq 0,
 \ee
 which means that the constraints $\chi_I$'s are of second class.
 Since the number of second class constraints is equal to the
 number of degrees of freedom, by imposing the constraints there remains no degree of freedom with respect to
 time $t$ in this theory. In this sense the theory is time independent. For quantization one should  use Dirac brackets
 which are defined  in terms of the canonical Poisson brackets  as follows:
 \be
 \{A,B\}_{\mbox{DB}}=\{A,B\}-\{A,\chi_I\}\chi^{IJ}\{\chi_J,B\},
 \label{Dirac}
 \ee
 where $\chi^{IJ}$ is the inverse of the matrix
 $\chi_{IJ}=\{\chi_I,\chi_J\}$.
 For example
 \bea
 \{\lambda^{-1},\bar\lambda^{-1}\}_{\mbox{DB}}&=&\frac{i}{2}e^K,\nn\\
 \{x^i,\lambda^{-1}(\bar\lambda^{-1})\}_{\mbox{DB}}&=&0,\nn\\
 \{\bar x^{\bar i},\lambda^{-1}(\bar\lambda^{-1})\}_{\mbox{DB}}&=&0,\nn\\
 \{x^i,\bar x^{\bar i}\}_{\mbox{DB}}&=&-\frac{i}{2}e^KG^{i\bar i},
 \label{DB1}
 \eea
 where in the last  equality above, $G^{i\bar i}$ is the inverse of the metric $G_{i\bar i}$.
 \par
 This is an example of non-commutativity of coordinates.
 Comparing the Dirac brackets obtained above with Poisson brackets (\ref{momenta1}) and the constraints
 (\ref{constraints}) it is easy to verify that in this theory, one can obtain true
 commutation relations (i.e. the Dirac brackets) by simply solving the momenta in terms of the coordinates by imposing the constraints $\chi_I=0$
 and inserting the solution into Poisson brackets (\ref{Poisson}). Of course one makes a mistake in evaluating the coefficient (here the factor 1/2)
 in doing so. What supports the validity of that wrong method (up to
 a factor 1/2 here) is the fact that by construction  by using the Dirac brackets one can safely assume that constraints are
 solved since using Eq.(\ref{Dirac}),
 \be
 \{A,\chi_I\}=\{\chi_J,B\}=0,\hspace{1cm}I,J=1,\cdots,2h+2.
 \label{vanishing}
 \ee
 and considering the {\em reduced phase space} possessing
 only coordinates  $\lambda^{-1}, x^i$ and their conjugate momenta
 $\pi,p_i$ with Dirac brackets,
 \bea
 \{x^i,p_j\}_{\mbox{DB}}&=&\frac{1}{2}\delta^i_j,\nn\\
 \{\lambda^{-1},\pi\}_{\mbox{DB}}&=&\frac{1}{2},
 \label{algebra}
 \eea
 and all other brackets vanishing. The above equations are consistent with the non-commutativity algebra (\ref{DB1}) and constraints
 (\ref{constraints}) as is expected from the identity (\ref{vanishing}). Finally we note that in constraint systems with second
 class constraints, for quantization the commutators are constructed in terms of the Dirac brackets instead of Poisson
 brackets. Therefore,
 \be
 p_i\to\frac{-i}{2}\frac{\rd}{\rd x^i},\hspace{1cm}\pi\to\frac{-i}{2}\frac{\rd}{\rd
 \lambda^{-1}}.
 \label{oper}
 \ee
 \subsection{The factor one-half of the action functional}\label{sec 3B}
 To obtain the holomorphic anomaly equations (\ref{holomorphic})
 we used the Poisson algebra (\ref{poisson}) but as is
 shown in section \ref{sec 3A}, the true algebra one has to use is
 the Dirac bracket algebra given by the Eq.(\ref{algebra}) which
 differs from the Poisson algebra by a factor one-half. Naively this makes a difference between the Schr\"{o}dinger equations (\ref{Schrodinger}) and
 the holomorphic anomaly equations in topological string theory. This problem can be  recovered by multiplying the right hand side of Eq.(\ref{action})
 by a factor one-half.

 Considering the action,
 \be
  S=\alpha\int_{CY_3\times R} C\wedge dC,
 \ee
  one should modify Eq.(\ref{momenta1}) slightly. For example,
 \be
 p_i=i\alpha e^{-K}G_{i\bar i}\bar x^{\bar i}, \hspace{1cm} \pi=-i\alpha e^{-K}\bar\lambda^{-1},
 \label{momenta2}
 \ee
 which are assumed to  satisfy the Poisson algebra (\ref{Poisson}). Although the constraints (\ref{constraints}) also
 modify correspondingly,  but one can show that the Dirac brackets (\ref{algebra}) do not
 modify. Consequently, still after modifying the action one has,
 \be
 \{x^i,p_j\}_{\mbox{DB}}=\frac{\delta^i_j}{2},\hspace{1cm}
 \{\lambda^{-1},\pi\}_{\mbox{DB}}=\frac{1}{2}.
 \ee
 The generators $H_i$ and $\bar H_{\bar i}$ given in Eq.(\ref{Hamiltonian}) should be modified   in order
 to obtain Eq.(\ref{c}) using the modified equations of motion,
 \bea
 {\cal O}_{,i}&=&\{{\cal O},H_i\}_{\mbox{DB}},\nn\\
 {\cal O}_{,\bar i}&=&\{{\cal O},\bar H_{\bar i}\}_{\mbox{DB}},
 \label{Heisenberg1}
 \eea
 The modified $H_i$, $\bar H_{\bar i}$ are,
 \bea
 H_i&=&-2\lambda^{-1} p_i-i\alpha C_{ijk}x^jx^k+\rd_iK[\pi,\lambda^{-1}]_++\Gamma^k_{ij}[x^j,p_k]_+
 +\rd_iK[x^j,p_j]_+,\nn\\
 \bar H_{\bar i}&=&-2G_{\bar i i}\pi x^i-\frac{i}{\alpha}e^{2K}\bar
 C_{\bar i\bar j\bar k}G^{\bar j j}G^{\bar k k}p_jp_k,
 \label{modified H}
 \eea
 One easily verifies that for  Schr\"{o}dinger equations (\ref{Schrodinger}) obtained by replacements (\ref{oper}) to be equivalent to
 the holomorphic anomalies (\ref{holomorphic}) one has to impose,
 \be
 \alpha=\frac{1}{2}.
 \ee
 Therefore the action functional for the quadratic field theory on
 $CY_3\times R$ that gives the holomorphic anomalies in topological
 string theory is that given in Eq.(\ref{trueaction}).
 
\end{document}